# A Framework for Next Generation Mobile and Wireless Networks Application Development using Hybrid Component Based Development Model

**Ahmed Barnawi, M. Rizwan Jameel Qureshi, and Asif Irshad Khan**

Faculty of Computing and Information Technology, King Abdul Aziz University1, Jeddah, Saudi Arabia

Email: ambarnawi@kau.edu.sa, rmuhammd@kau.edu.sa, aikhan@kau.edu.sa

**Abstract** – The IP Multimedia Subsystems (IMS) that features in Next Generation Networks (NGN) offers the application developer (third party) abilities to map out applications over mobile telecommunication infrastructure. The IMS comes about with APIs useful for mobile application developers to create applications to meet end-users' demands and comply with the provider's infrastructure set up at the same time. Session Initiation Protocol (SIP) is a signaling protocol for this architecture. It is used for establishing sessions in IP network, making it an ideal candidate for supporting terminal mobility in to deliver the services with improved Quality of Services (QOS). The realization of IMS's virtues as far as software design is concerned is faced by lack of standardizations and methodologies throughout application development process. In this paper, we report on progress on ongoing research by our group toward putting together a platform as a testbed used for NGN application development. We examine a novel component based development model used for SIP based mobile applications. The developed model is to be used as framework for general purpose application development over the testbed. We apply this model on MObile Mass EXamination (MOMEX) system that is an application attracting the interest of educational authorities around the world due to its potential convenience.

## 1. Introduction

Next Generation Mobile and Wireless Networks (NGMWNs) plays central rule to bring together the users, operators, and application developers. From an end-user's point of view, NGN should consistently present him/her with the same personalized services whatever the network that serves him/her and whatever the terminal technology he/she uses.

From applications developers' point of view, NGN offers resources and interfaces Application Programming Interfaces "APIs" through which the applications are deployed. From an operators' point of view, NGN is concerned with provisioning and guaranteeing end to end (e2e) Quality of Service (QoS) in the context of all-IP heterogeneous network. In order to meet with such high expectation, NGN network have gone through many changes to come up optimized network design without incurring major changes to network infrastructure.

IP Multimedia Subsystem (IMS) is considered as the cornerstone for NGN. IMS is best described as the glue between the "global" applications world (Internet) and the mobile world. The IMS was designed to make it easy for third party developers to deploy their applications over mobile networks. According to the standards, IMS is defined in the form of reference architecture to enable delivery of next-generation communication services of voice, data, video, wireless, and mobility over an Internet Protocol (IP) network [1].

Fixed Mobile Convergence (FMC) is technological junction of fixed and mobile services. FMC offers a unique way to connect mobile phone to a fixed line infrastructure so that the operators can offer services to their users irrespective of their access technology, location and terminal. FMC Uses IMS to provide innovative, reliable new services. In FMC, the same handset can access the services through a fixed network in addition to a wireless network. It can be used in home or office and also while traveling [2].

The complexity of today's data communications networks necessitates complete, realistic and sophisticated testing playground for verifying and validating their functionalities. Testing in production and commercial networks is typically forbidden since they present a high degree of risk factors for service availability. Therefore, the optimum objective for our research group is the implementation of NGN testbed to be used as a platform for next generation mobile application development.

Signaling in IMS network is based on a Session Initiation Protocol (SIP). The SIP based architecture provides a multiservice environment with multimedia capabilities. IMS contains Home Subscriber Server (HSS), which is the central storage area for user-related information such as his/her security related information or the service to which the user is subscribed to. It is also consists of the Serving Call Session Control Function (S-CSCF) which acts as the central node of the signaling plane. S-CSCF on one hand is connected to the Application Server that hosts the application and on the other it is connected to HSS and the mobile IMS either through the

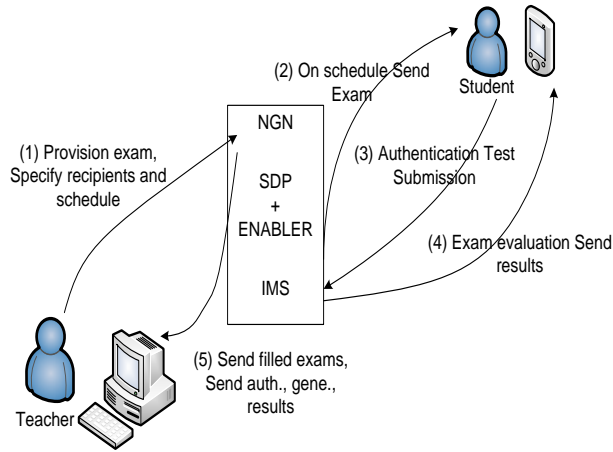

Figure 1- IMS- based mobile exam scenario

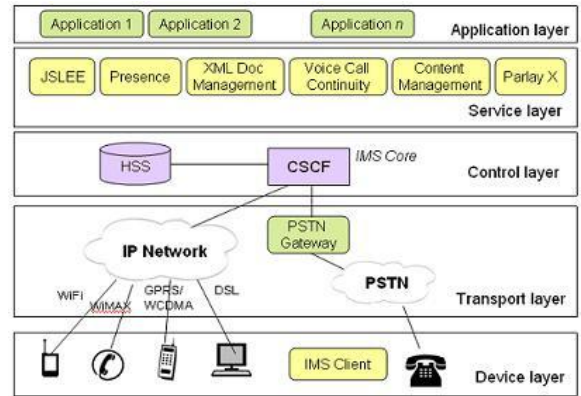

Figure 2- IMS architectural diagram Source [3]

Proxy CSCF (P-SCSF) if the client resides in its own area of serving or Interrogator CSCF (I-SCSF) if a client is being served by another S-CSCF.

Although IMS based infrastructure has matured enough for commercial deployment still development of SIP based applications lacks standardized software engineering methodologies. In this paper, we are proposing a framework for NGN mobile application development. This framework aims to facilitate creation of mobile applications on top of ubiquitous infrastructure.

To achieve this goal, we are utilizing the use of our developed NGN testbed that simulate IMS infrastructure. Such framework will facilitate managing process of software development and testing prior to actual run over operator's infrastructure. We are applying the developed framework on creation of a mobile application. The application developed is MObile Mass EXamination (MOMEX) System.
MOMEX System expedites the examination process by automating various activities in an examination such as exam paper setting, scheduling and allocating examination time and evaluation etc.

The MOMEX system will assess to students by conducting mobile based objective exam. This will be highly customizable for any university who acquired to adopt similar IMS based examination system and faculties to create their own dashboard (create set of questions, creates groups, adds related students into the groups, schedule exams, etc.). Further, the exams will be associated with specific groups so that only associated students can appear for the test, result will be notified to the student either through SMS/email as shown in Figure 1.

In software development phase, MOMEX follows Component based Software Development (CBD) life cycle, CBD model encourage assemblies of components for systems development instead of building everything from scratch also, CBD support the development of components as reusable entities.

The paper is organized as: section 2 describes IMS architecture, API details and CBD Model, section 3 describe related works and section 4 provides the proposed solution to the research problem in hand.

## 2. IMS Architecture

IMS is the 3rd Generation Partnership Project's (3GPP) vision for a converged telecommunications architecture that merges cellular and Internet technologies to uniformly deliver voice, video, and data on a single network with a better quality of service (QoS). IMS is designed with building blocks, enabling telecommunication operators to deliver new services in a more flexible way. IMS network architecture is composed of three main layers that are transport, control and service layers, which are shown in figure 2. The separation of layers makes it easy to standardize the interfaces and interconnect the systems [3]. 3$^{rd}$ Generation Partnership Project (3GPP) defines IMS as an architecture framework for delivering multimedia services for both wired and wireless based technologies based on the Internet Protocol (IP) [1]. SIP has emerged as a vital technology for controlling communication in IP-based Networks.

*2.1 Main layers of IMS Architecture*

Transport layer is designed to provide network access. Initiation and termination of SIP sessions are done by this layer. IMS devices connect to an IP network in the transport layer using different technique including Cable, Wireless Fidelity (WiFi), digital subscriber link (DSL), Session Initial Protocol (SIP), General Packet Radio Service (GPRS), and Wideband Code Division Multiple Access (WCDMA) etc.
Control layer mainly contains Call Session Control Function (CSCF) and Home Subscriber Server (HSS). Routing and generating billing are the main responsibility of this layer.
Application layer allows service providers to offers different multimedia services to its users. The application servers are hosting/executing the services and provide interface against the control layers using the SIP protocol. Presence server, Instant Messaging Server, Group List Management Server etc are some core application servers.

Several international telecommunication and Internet bodies are responsible for the standardization of IMS, SIP and Java technologies like 3rd Generation Partnership Project (3GPP), European Telecommunications Standard Institute (ETSI), Internet Engineering Task Force (IETF), Java Community Process (JCP) etc. JCP is also involved in the standardization of Java APIs to facilitate the ease and fast deployment of IP services using SIP. The main IMS system consists of different software components inter-operating to provide services and management to the network. Following are some of the software components that support to IMS.

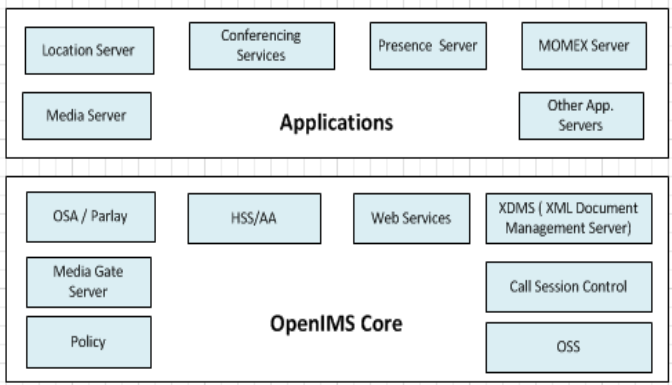

Figure 3- OpenIMS testbed @ FOKUS [4]

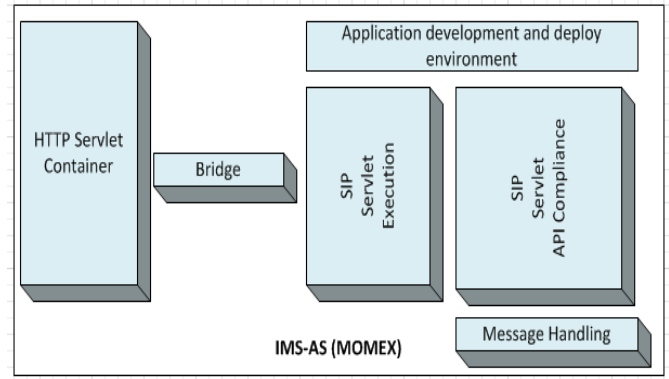

Figure 4- SIP Servlet Application Server [6]

## 2.2 IMS API Details

The proposed IMS testbed is based on FOKUS Open IMS Playground. FOKUS (IMS Core) used application programming interface (APIs) to support component based development (CBD) to get a highly customizable Mobile Examination system using Java Bean component model for the object-oriented programming language. JavaBeans add a standard way to manipulate Java classes conforming to the Beans specification within an integrated development environment (IDEs). Simple object access protocol (SOAP) based Network Abstraction API, Parlay X, XML (Extensible Markup Language), Document Management Server, Fokus myMONSTER TCS kit, Home Subscriber Server (HSS), FHoSS and Mysql are some set of powerful simple and highly abstracted building block of telecommunication capabilities supporting.

Main development platform is Linux & programming languages are C and Java. Figure 3 shows FOKUS open IMS testbed [4]. Main APIs to develop IMS application are as follows.

IMS Core component- This module APIs comprises all the programmatic interfaces exposed by the IMS Core in order to develop IMS applications.

The Open IMS CSCFs (Proxy, Interrogating, and Serving) Component [5]- It uses SIP express router (SER) which works as SIP registrar, proxy or redirect server and is able of handle many thousands of calls per second. Providers can extend the functionality of SER Platform by defining new components and incorporate them with the SIP Express platform with the availability of Plug and Play interface in FOKUS Open IMS Playground. Open IMS CSCFs consists of four main components.

Proxy-CSCF Module (PCSCF) API is supposed to provide the functionality required for a Proxy-CSCF. Interrogating-CSCF Module (ICSCF) API is providing the functionality required for an interrogating-CSCF. While the Serving-CSCF Module (SCSCF) API is supposed to provide the functionality required for a serving-CSCF, and the SIP-to-IMS Gateway Module (SIP2IMS) This API module is supposed to provide the translation capabilities to enable normal SIP User Endpoints to use the Open IMS Core through the Proxy-CSCF. The IMS Service Control API (ISC) is supposed to provide support for the ISC interface between the Serving-CSCF and the Application Servers.

FOKUS Home Subscriber Server (FHoSS) Component [7] is a prototype of Home Subscriber Server (HSS), which is entirely written in Java and is based on open source MySQL database. FHoSS stores the user files via Sh interface (de.fhg.fokus.sh package API) point to Application Servers, and it communicates with CSCFs via Cx interface (de.fhg.fokus.cx package API) besides, it connects with Security Domain via Zh interface using (de.fhg.fokus.zh) package API. For each interface there is an implementation which can be found in the de.fhg.fokus.hss.server and subsequent packages. JavaDiameterPeer Package is available for TCP and Diameter base protocol supports, it is designed to offer best performance while fulfilling IMS requirements. All the above support is under FHoSS Package [7].

FOKUS created a Servlet container SIP Application Server (SIP AS) [5] in IMS environment called SIPSEE (SIP Servlet execution environment). SIPSEE is based on SIP Servlet API 1.0 (JSR 116) that is an appropriate environment to deploy SIP servlet applications building upon a JAIN SIP stack. Main components of the SIP AS consist of SIP Servlet Execution, SIP Message Handling, Application Deploy Environment, SIP Servlet API Compliance and Bridge components are shown in figure 4.

SIP- AS based on Parlay X [5] - FOKUS offers API that works on top of a HTTP- and a SIP Servlet Container enabling distributed Web Service Clients or other Web Service Servers to take advantage of this converged SIP Servlet Application Server. New and innovative applications can be quickly comprehended and generated using Parlay X.

FOKUS myMONSTER Telco Communicator Suite (TCS) [7] - IMS client can be easy plug-and-play on multiple platforms like Windows Mobile, Linux, Windows & Google's Android, TCS offers developers an environment for development of variety of internet and telecom services including telephony, chat, messaging, social networking and contact manager.

TCS framework gives developers the flexibility and options to extend framework with their own components as shown in Figure 5.

The IMS engine interface is found under the packages javax.ims, javax.ims.core and javax. ims.core.media API packages, de.fhg.fokus.monster.apps.impl.ims package contains collection of classes and interfaces that are useful for communication functions on top of IMS, while the package de.fhg.fokus.monster.apps.impl.ims.session is a collection of classes and interfaces for session management. API reference for application interface (call, chat, group, location, messaging, presence, contacts, etc) definitions is available under packages de.fhg.fokus.monster.apps.

Component-based software engineering (CBSE) is used to develop/assemble software from existing components. CBD technologies comprised of implementing a component into a system through its well defined interface [9]. Whereas, the organization of quality features of a component is not assisted by CBD technologies.

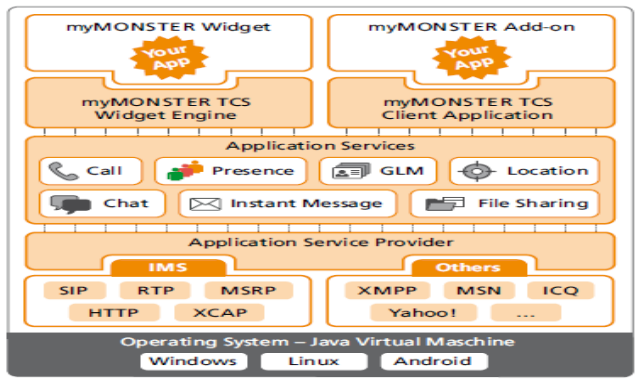

Figure 5- FOKUS myMONSTER framework [7, 8]

*2.3 Component Based Software Development*

The areas where such technologies are extensively practiced, the quality features are neither of main concern nor adequately tackled.

It is a well-known truth that CBD is important for large and complex systems but why it is important for mobile device applications. It tackles vital concerns such as productivity, high level of abstraction, partitioning of the system development process from the component development process and reusability [10]. Reusability offers a number of advantages to a software development team [11]. An assembly of component assembly leads to a 70 percent reduction in development cycle time; an 84% reduction in project cost, and a productivity index of 26.2, compared to an industry norm of 16.9 [11]. For the development of mass mobile examination system, CBD is a smart method, but due to its explicit requirements such as real time, safety, reliability, minimum memory and CPU consumption, standard component models cannot be used [10]. Rather than, a new CBD methodology is very much needed for the development of mobile mass examination system to deal with its specific requirements.

MOMEX software development model is based on component based software development. One of the principles of computer science field to solve a problem is divide and conquer i.e., divide the bigger problem into smaller chunks. This principle fits into component based development. The aim is to build large computer systems from small pieces called a component that has already been built instead of building complete system from scratch.

Reuse of software components concept has been taken from manufacturing industry and civil engineering field [12]. Manufacturing of vehicles from parts and construction of buildings from bricks are the examples. Car manufacturers would have not been so successful if they had not used standardized parts/components. Software companies have used the same concept to develop software in standardized parts/components. Software components are shipped with the libraries available with software.

Microsoft Corporation and Sun Microsystems are two major software-providing organizations. These companies have provided components with their software to market themselves successful and their tools are widely used in software industry. Most of the offered tools provide an IDE (Integrated Development Environment). IDE provides an environment in which components are available in the toolbox or in the reference library like a car assembling plant. We do not need to develop the components during the assembling of the car but they are there and we timely assemble them. Similarly in IDE, the standard components such as text box, label box and command button are available in the toolbox and we just integrate and use them [12].

## 3. Related Works

Many research bodies/organizations are developing better ways to manage exams systems and assessments.
[13] looked Web-based Exam Management Systems (EMS) as an effective solution for mass education evaluation. They proposed a web based system to conduct online exam activities such as registration, online exam, auto grading etc. the authors concluded that the presented system saves instructors from suffering and tedious of grading works, further in the paper the authors pointed out that 94 % of the students like the user interface and 85 % agree that the system is usable [13].

Web-based examination system is an effective tool for mass education evaluation [14]. A novel online examination system based on a Browser/Server framework is proposed. Exam and auto-grading for objective questions and subjective questions has been successfully applied to the distance learning evaluation of basic operating skills of computer science, courses for the high school graduates in Zhejiang Province, China[14].

The author [15] described a distributed cross-platform examination system based on web service and COM components using visual C #. [16] Used XML and COM components, and combined Browser/Server model (B/S) and Client/Server model (C/S) to solve the shortcoming of existing examination systems.

[16] used the random linear algorithm for question selection and intelligent examination paper grouping algorithm based on question bank structure, examination question structure and controlling parameters have been constructed and realized. They [16] developed a system in Java 2 Enterprise Edition (J2EE) environment and an object oriented software engineering model, UML, Java and Extensible Markup Language (XML) techniques are used in the system development successfully implementation to twenty seven universities or colleges and gained better economic and social benefit[16]. The authors deemed it appropriate to mention that no research has been done to propose a novel model for SIP based mobile based examination systems as per their knowledge. This objective of this research is to propose a novel CBD model for the development of IMS application.

The research problem is: "How to propose an improved systematic component-based development process model for development of IMS-based mass examination system?".

The CBD model proposed [17] was not that comprehensive. In fact its main focus was on reuse activities and it lacked in domain engineering activities. The domain engineering activities are vital to populate the repository to reuse components in current and future software projects. The said limitation in the CBD model (of John and Victor) motivates the author to propose a new CBD model that includes both CBD and domain engineering activities.

[18] proposed another CBD model that was a modified version of Waterfall model. It was a time consuming and costly model due to verification of phases like Waterfall model. This model emphasized on freezing the requirements. That is why; it was not suitable for commercial projects where requirements of the project development are dealt dynamically. The model was suitable only when comprehensive or stable requirements were provided from the very beginning. The shortcomings of CBD model [18] become source of inspiration to propose a new CBD model that will deal with run time requirements dynamically to meet the needs of commercial software projects.

An incremental model was proposed for CBD by the author [19]. Analysis and construction were only two phases in the incremental model. It was neither that comprehensive nor properly covering system development life cycle phases for CBD software projects. The usage of specific case tool to adopt the model was another limitation. The limitations of their model encourage the author to propose a new CBD model. The new CBD model will have relatively more comprehensive system development life cycle phases for both CBD and domain engineering activities. Moreover, there will be no constraint to use any specific case tool while using the proposed model to develop CBD software projects.

A four-stage component-based development process model was proposed by the authors [20]. They introduced a new technique to integrate off-the-shelf components with the newly developed components in the then existing CBD process models. However, they paid less rather no attention to the integration of internally developed components. The same motivates the author to propose a new CBD model.

The author [21] used repository on the design phase. The design phase is not that appropriate phase to use repository. This is because; using repository at analysis phase helps a software engineering team to identify new and reusable components. Moreover, it has many benefits such as; initiation of CBD and domain engineering activities and more accurate estimation of cost, schedule and effort to develop and integrate components in a CBD project. The proposed CBD model will use repository at analysis phase to avail the above mentioned benefits.

It is with this logic and background in view that the author feels motivated to propose a new CBD model. The same is accomplished by proposing a new CBD model as a solution to the research problem as follows.

## 4. The Proposed CBD Model for the IMS

A new component-based development (CBD) model has been proposed for an IMS-based mass mobile examination system as a solution for the research problem. A CBD model is a process model that provides a framework to develop software from previously developed components. The main phases of the improved CBD are 'Project Planning', 'Analysis', 'Adaptation, Engineering & Integration' and 'Testing'. An improved CBD model to be proposed is shown, in figure 6.

**Project Planning Phase**- A customer is communicated at the start of the project to gather basic requirements. Initial use cases are developed at this stage to prepare project during the 'Project Planning' phase. Project specification or proposal document is composed of feasibility and risk assessments that are performed to prepare a cost benefits analysis (CBA) sheet. CBA sheet helps to estimate whether the software project is feasible for the customer or not. Use case diagrams of a system are helpful in understanding functional requirements of the system and user's interaction with system. Main aim of SIP based Mobile Examination System (MOMEX) is to facilitate mobile based examination service support to the university in a timely and reliable manner. System mainly targets students and it enables interactive way of handling examination system which includes registration, user management, report generation, question paper matrix entry, question pool management, data entry and examination conduction etc, done in an integrated and centralized way. Students access the system using mobile interface. It lets a student to view list of tasks assigned to them. In particular, schedule of the exam, appearing in exam, view result etc. Web interface is oriented to faculty staff and administrator of the System for generating support services. Figure 7.1 to Figure 7.4 shows a unified modeling language (UML) use case diagram for MOMEX prototype. As shown in figure, there are three primary actors in the MOMEX system.

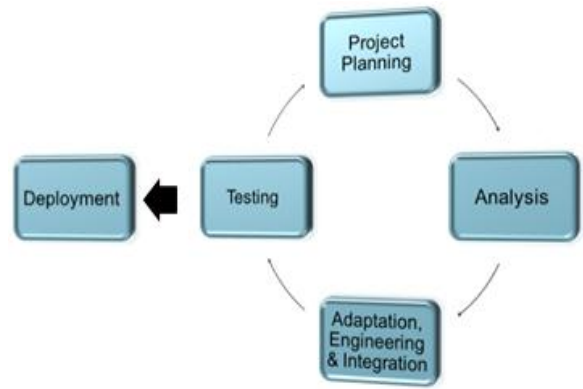

Figure 6- The Proposed New Component-Based Development Process Model

For Administrator actor, MOMEX provides ability to configure the System, such as create profile for user with their access role, setup courses to conduct exams etc, For Faculty actor as shown in figure 7.1, MOMEX provide ability to work as question bank settler (set questions for allowed courses in question bank) also, faculty can act as exam scheduler (schedule and set a exam for assigned courses, register students to groups, send notifications etc).

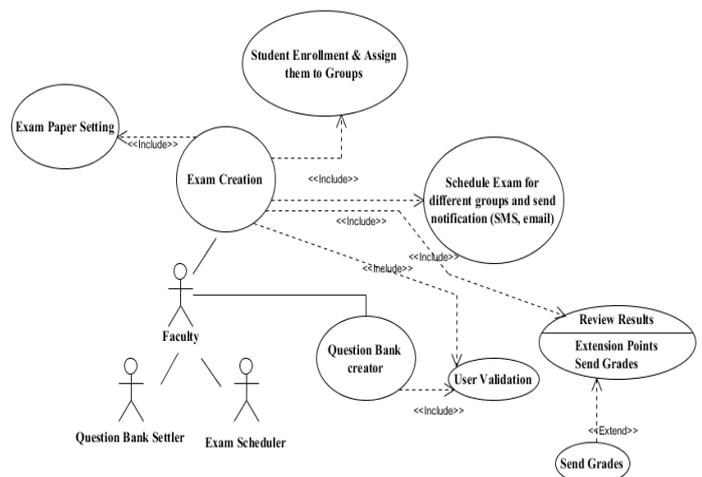

Figure 7.1 - Use case diagram of faculty

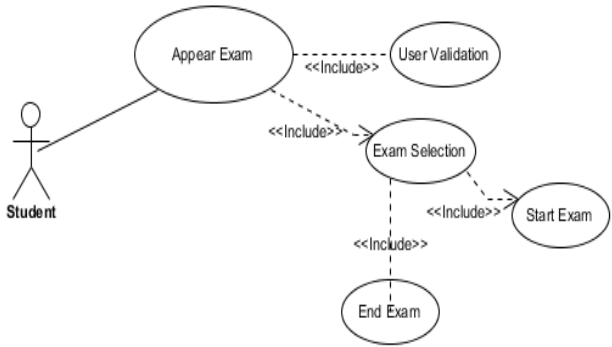

Figure 7.2 - Use case diagram of student

For Student actor as shown in figure 7.2, MOMEX provide ability to appear in an exam as per their exam schedule. Send queries related to exam, while System Administrator and System as shown in figure 7.3 and 7.4 respectively, provides various supporting functions to run application smoothly like automatically sends reminder for exams to registered students, save exam results in database, conduct exam, Generate and send result to students etc, all users (faculty, Administrator, Students) have to undergo user validation before using their dash board (system functionalities).

**Analysis' Phase**- Analysis phase is initiated if the customer approves the proposal. This is the phase where an analyst gathers the detailed requirements of the system to be developed. A domain analysis is performed to find a suitable architecture for the application to be developed [21]. An architectural model of application is developed that enables software engineer to:
- evaluates efficiency of design;
- judge options of design;
- minimize potential threats coupled with software development [10]

'Analysis, Selection & Risk Management' is the phase where an analyst tries to identify and select those components that can be reused from the components repository. The selection of reusable components is important to improve productivity of component-based software development. Risks about new and existing components are also evaluated and managed. Software engineering methods are applied to develop new components for those requirements which cannot be fulfilled from already developed components.

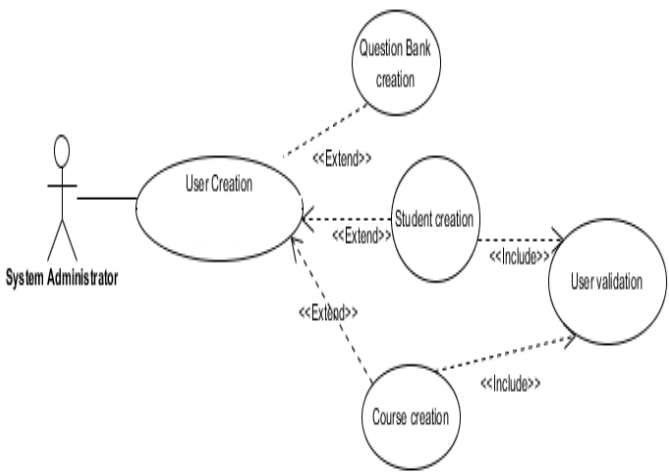

Figure 7.3 - Use case diagram of system administrator

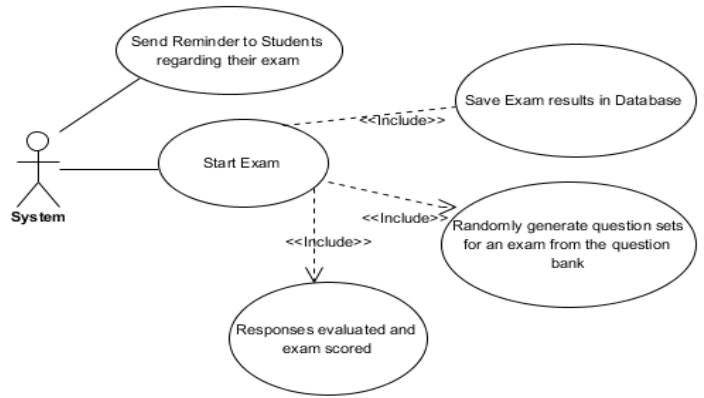

Figure 7.4 - Use case diagram of system

Reusable components need qualification, adaptation and composition. Component qualification makes sure that the selected component will execute the desired functionality, integrate easily into the structural design of new application and demonstrate the quality attributes (e.g., reliability, performance, usability) [21]. The properties, behavior and relationships among components are identified. Core objective of this phase is to reuse components as maximum as possible, rather than reinventing the wheel. It will also improve productivity and efficiency of software engineers.

The processes of student appear in an examination is shown using sequence diagram figure 8.

Client (Student) will interact with IMS network via his mobile, Logger interface create session if successfully validated by IMS network, connectivity Application Programming Interfaces (APIs) process requests for client provide access to applications (test), once student activate his exam session, questions will be randomly picked from Exam pool Object. Once exam is over or exam time expired, result auto marked by system and send to student.

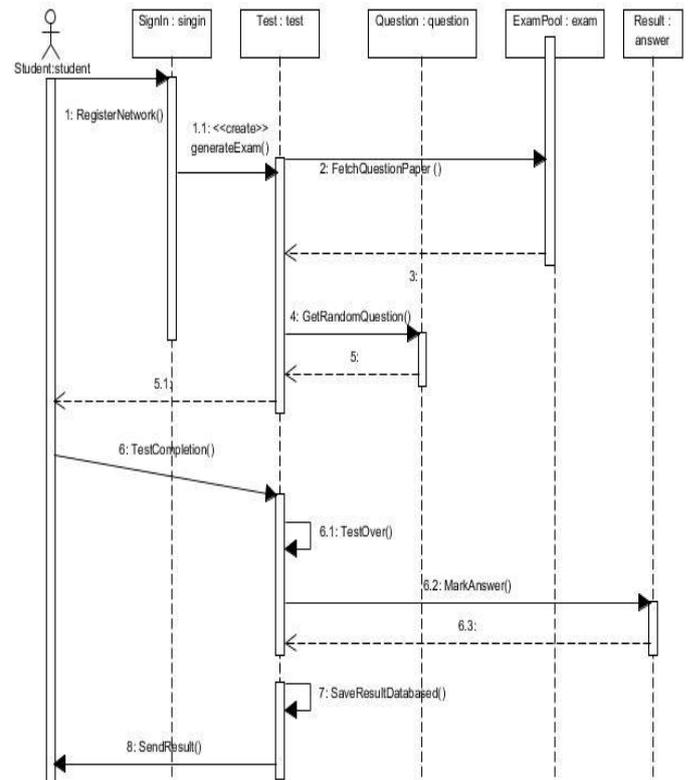

Figure 8- Student exam scenarios.

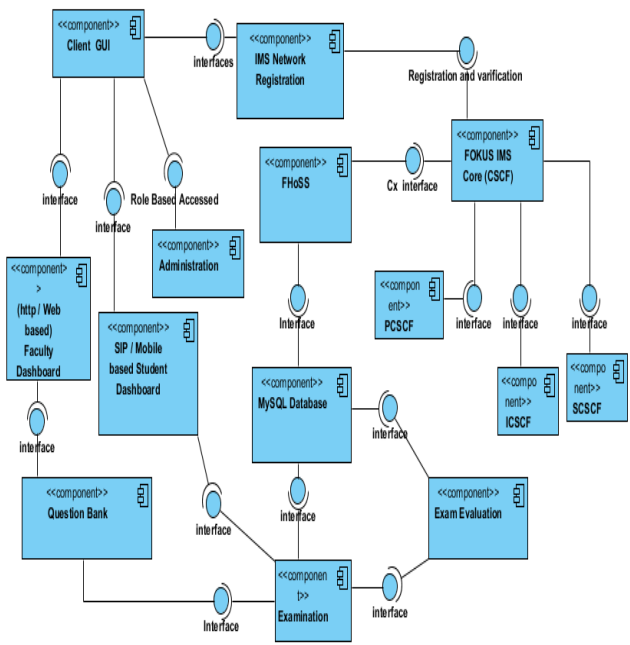

Figure 9- UML Component Diagram of the System

Detail description of the components: Architecture developed within the MOMEX project is based on the specialized CBD. Each component has its own purpose. Several rules and practices of CBD are applied in this project. MOMEX mainly contains the following main components (modules) as shown in figure 9.

Network Registration Component: Client first need first registered in the network, Registration component is responsible for Network registration and validation of client to IMS platform. In particular, user registration procedure deals with mutual authentication between the user equipment (UE) and the IMS platform by using the 3GPP Authentication and Key Agreement (3GPP-AKA) mechanism. This process is carried out by means of IMS core infrastructure (IMS enabler) including control functions and subscriber database.

Administration Components: This module is responsible for creating different users of the system with their assigned roles. Creation of different courses which are the building block for creation of question bank and the test are other responsibilities of this module.

Question Bank Component: This component interacts with Exam creation component for creation of question bank for a particular course.

View Component: Is responsible for generating GUI (graphical user interface) for different user based on their access role, along with its attributes.

Exam Creational Component: Faculties use their dashboard to create questionnaire (fixed question exam or random question exam) by assembling Question Bank Components resulting questionnaire contains active code and is pluggable into the dashboard of the students. The faculty needs only to drag and drop questions within an IDE in a questionnaire container. The connections for the interactions between the questions and the questionnaire are done automatically. The resulting questionnaire and questions implement all needed interfaces to be plugged into the student application and for the final evaluation by a master copy of the questionnaire together with the right answers.

Examination Component: This component is responsible for conducting mobile based examination, generating questions, storing the attended answers in the database.

Exam Evaluation Component: Student attempted questions can be evaluated automatically by faculty using evaluation component. Faculty uses their dash board to evaluate attempted returned answer made by students.

This module holds all questionnaires and offers for automatically evaluable questions an evaluator bean, which is connected with master copy of questionnaire as obtained from design phase, which contains the correct answers. Evaluation for each questionnaire is passed to a report generator bean. Report can be edited manually by faculty to include corrections of not automatically evaluated answers, comments and final grade of exam.

**Adaptation, Engineering and Integration' Phase**- The reusable components are customized according to the requirements of the new system to be developed and tested. The next part is integration and testing again. A common component adaptation technique called component wrapping is used if programmer is using black box components.

Model-view-controller (MVC) is used for design pattern as MVC isolate business logic from the user interface which is the requirement of the system. In MVC, Model represents information (data) of the application and business rules used to manipulate data, View corresponds to elements of user interface such as textbox, checkbox items, etc, and Controller manages details involving communication between model and view. The controller handles user actions such as keystrokes and mouse movements and pipes them into model or view as required [22].

**Testing Phase**- The new components are designed, developed and tested on unit basis. Integration and system tests of the newly developed and of the reused components are performed. A customer is requested to evaluate and verify software, whether it meets his/her requirements or not during the testing phase. The software is ready to deploy at customer site.

## 5. Conclusion and Future Work

An ongoing research is presented in this paper by putting together a platform as a testbed for NGN application development. We propose a novel component based development model used for this SIP based mobile applications. The proposed model used as a framework for general purpose application development over the testbed. The objectives of IMS based Mobile examination System approaches are explained with reasons and advantages identified. Main components of IMS service architecture with their roles are also described. The approach leads to a highly modular and extensible integrated system. Future work is to validate novel component based development model using a case study of IMS testbed. Multi-tier applications architecture (client, web, and business) is adopted, as per the needs of case study i.e., MVC design pattern [23].

## 6. Acknowledgement
The authors would like to thank King Abdulaziz City for Science and Technology (KACST), Saudi Arabia for funding this ongoing research project.

**References**

[1] Barnawi, A., Akkari, N., Emran, M., Khan, A.I. "Deploying SIP-based Mobile Exam Application onto Next Generation Network testbed", IEEE *Xplore* Digital Library, first Saudi Int. Electronics, Communications and Electronics Conf. (SIECPC'11), Riyadh, KSA, 2011, pp. 1–6.

[2] Techwriters Future,"Fixed Mobile Convergence", [Online], viewed December 20, 2011, available: by http://ipv6.com/articles/mobile/Fixed-Mobile-Convergence.htm

[3] Sike Huang, "Mobile Telemedicine System based on IMS/SIP platform", Master Thesis, School of Info. and Communication Tech., Royal Inst. of Tech., Stockholm, Sweden 2009 ECS/ICT-2009-27.

[4] Magedanz, T. Witaszek and D. Knuettel, K. "The IMS playground @ FOKUS-an open testbed for generation network multimedia services", IEEE Computer Society Washington, DC, USA, First Int. Conf., Tridentcom, pp. 2 – 11, Feb. 2005.

[5] Fraunhofer Inst. for Open Communication Systems FOKUS, "OPEN IMS PLAYGROUND, Open Testbed for IMS Tech.", [Online], viewed October 01, 2011, available: http://www.fokus.fraunhofer.de/en/fokus_testbeds/open_ims_playground/components/index.html

[6] M. Sher, S. Wu and T. Magedanz "Security Threats and Solutions for Application Server of IP Multimedia Subsystem (IMS-AS)", Workshop MonAM, Tübingen, Germany September, 2006.

[7] Fraunhofer Inst. for Open Communication Systems FOKUS, "The myMONSTER Telco Communicator Suite (TCS) enables", [Online], viewed September 10, 2011, available: http://www.fokus.fraunhofer.de/en/fokus_testbeds/open_soa_telco_playground/_docs/TCS_product_broschure_2010-10.pdf

[8] Fraunhofer Inst. for Open Communication Systems FOKUS, "Fraunhofer FOKUS Competence Center NGNI ", [Online], viewed September 13, 2011, available: http://www.openepc.net/about_us/fokus_ngni/index.html

[9] H. Hansson, M. Åkerholm, I. Crnkovic and M. Törngren, "SaveCCM – a Component Model for Safety-Critical Real-Time Systems", in Proc. of the 30th EUROMICRO Conf. (EUROMICRO'04), France, 2004.

[10] Software Engineering, 6th ed. McGraw-Hill Co., Pressman, R. S., 2005

[11] M. Rizwan Jameel Qureshi., "Reuse and Component Based Development," in Proc. of Int. Conf. Software Engineering Research and Practice (SERP'06 Las Vegas, USA), pp. 146-150, 26-29 June 2006.

[12] M. Rizwan Jameel Qureshi and S.A Hussain., Reusable Software Component-Based Development Process Model", Advances in Engineering Software, Elsevier Ltd, Amsterdam, the Netherlands. Vol. 39/2, pp. 88-94, February 2008.

[13] "Magdi Z. R., Mahmoud S. K., Ahmed E. H., and Mahmoud A. Z." An Arabic Web-Based Exam Management System, Int. J. of Electrical & Computer Sciences IJECS-IJENS Vol: 10 No: 01, pp. 48-55,2010.

[14] Yuan Zhenming, Zhang Liang and Zhan Guohua, "A Novel Web-Based Online Examination System for Comp. Science Education", 33rd ASEE/IEEE Frontiers in Education Conf., S3F-7-S3F-10, 2003.

[15] Li Jun "Design of Online Examination System Based on Web Service and COM", the 1st Int. conf. on Information Science and Engineering (ICISE2009), 2009

[16] Wang Aimin and Wang Jipeng "Design and Implementation of Web-Based Intelligent Examination System", in Proc. of the 2009 WRI World Congress on Software Engineering, IEEE Computer Society Washington, DC, USA - Volume 03, 2009.

[17] John, W. B. and Victor, R. B. The Software-Cycle Model for Reengineering and Reuse. Proc. of conf. on TRI-Ada '91: today's accomplishments; tomorrow's expectations, USA, 1991, pp. 267-28.

[18] J.Q. Ning, "A Component-Based Software Development Model", in Proc. Of the 20th Conf. Computer Software and Applications, Seoul, Korea, pp. 389-394, 1996.

[19] E.S. de Almeida, A. Alvaro, D. Lucredio, A.F. do Prado and L.C. Trevelin, "Distributed Component-Based Software Development: An Incremental Approach", in Proc. of the 28th Annual Int. Conf. Computer Software and Applications, pp. 4-9, 2004.

[20] J. Hutchinson, G. Kotonya, I. Sommerville and S. Hall, "A Service Model for Component-Based Development," in proc. Of the 30th EUROMICRO Conf., Rennes- France, pp. 162-169, 2004.

[21] Mark Lycett, George M. Giaglis. "Component based information systems: Towards a framework for evaluation", in proc. of the 33rd annual int. conf. on system sciences, Hawaii, 4–7 January 2001.

[22] rj45, "Simple Example of MVC (Model View Controller) Design Pattern for Abstraction", Online], viewed September 10, 2011, available: http://www.codeproject.com/KB/tips/ModelViewController.aspx

[23] Krasner, G.E, and Pope,T.S. , "A Description of the Model-View-Controller User Interface Paradigm in the Smalltalk 80 System", J. of Object Oriented Programming 1(3), pp. 26-49. 1988.

[24] Brereton, P. and Budgen, D., "Component-based systems: A Classification of Issues", Computer J., 33, pp. 54-62, 2000.

[25] Dogru, A.H. and Tanik, M.M., "A Process Model for Component-Oriented Software Engineering:, IEEE Software, 20, pp. 34-41, 2003.

[26] Crnkovic, I. and Larsoon, M., "Building Reliable Component-Based Software Systems", Artech House, London, 2002.

[27] E.S. de Almeida, A. Alvaro, D. Lucredio, V.C. Garcia and S.R. de Lemos Meira, "A survey on software reuse processes," IEEE Int. Conf. Information Reuse and Integration, pp. 66-71, 2005.

[28] G. Kotonya, I. Sommerville and H. Steve, "Towards A Classification Model for Component-Based Software Engineering Research," in Proc. 29th Conf. on EUROMICRO, pp. 43, 2003.